# Electromagnetically induced absorption in metastable $^{83}$Kr atoms

Y. B. Kale*, V. B. Tiwari, S. R. Mishra, S. Singh and H. S. Rawat

Laser Physics Applications Section, Raja Rammanna Centre for Advanced Technology, Indore, - 452013, India.
*Corresponding author: yogeshwar.kale@gmail.com

**We report electromagnetically induced absorption (EIA) resonances of sub-natural linewidth (FWHM) in metastable noble gas $^{83}$Kr* atoms using degenerate two level schemes (DTLSs). This is the first observation of EIA effect in a metastable noble gas atoms. Using these spectrally narrow EIA signals obtained corresponding to the closed hyperfine transition from $4p^55s[3/2]_2$ to $4p^55p[5/2]_3$ hyperfine manifolds of $^{83}$Kr* atoms, we have measured the Landé's g-factor ($g_F$) for the lower level (F = 13/2) of the closed transition accurately with small applied magnetic fields of few Gauss.**

http://dx.doi.org/

Quantum interference in the evolution of atomic states during light-atom interaction results in many interesting phenomena including electromagnetically induced transparency (EIT) and electromagnetically induced absorption (EIA) [1]. Opposite to EIT effect, the phenomenon of EIA refers to the enhancement in the absorption of a probe laser in presence of a strong pump laser light interacting coherently with the atoms [1, 2]. The EIA effect in a degenerate two level scheme (DTLS) has been widely studied [3-13] and has possible applications in optical switching, information storage and sensitive magnetometry. In the DTLS, lower and upper levels may have multiple Zeeman sub-levels where a strong pump (control) laser beam modifies the absorption of a weak probe laser beam of the same frequency. The Zeeman sub-levels of either levels (lower or upper) involved in the transition play an crucial role in the realization of EIA in these DTLSs. If the pump and probe have same polarizations, the transfer of population (TOP) among Zeeman sub-levels of lower level plays important role in realizatrion of EIA. On the other hand, if the pump and probe have different polarizations, the transfer of coherence (TOC) leads to EIA. An anomalous EIA observation [5], which can not be explained by either of the above two mechanisms (i.e. TOP and TOC), has been explained using dressed-atom multiphoton spectroscopy (DAMS) model [11]. The EIA has been investigated in different alkali atoms under different conditions [3, 6, 10].

In this letter, we report EIA resonances of sub-natural linewidth (FWHM) in metastable $^{83}$Kr ($^{83}$Kr*) gas atoms. As per our knowledge, this is the first observation of EIA in metastable noble gas atoms. In contrast to the alkali atoms, the $^{83}$Kr* atom offers multiple DTLSs to be investigated due to manifold hyperfine structure in lower and upper levels. Here, EIA and EIT both, have been observed for the DTLSs involving the transition from $4p^55s[3/2]_2$ to $4p^55p[5/2]_3$ manifolds of $^{83}$Kr* atom. The effect of control laser power on the EIA signal for a closed transition (i.e. F = 13/2 to F' = 15/2) has been studied. Using the narrow EIA signals for the closed transition, the $g_F$-factor of the lower hyperfine energy level F = 13/2 has been measured accurately at low value (few Gauss) of applied magnetic field. The accurately measured values of the $g_F$-factor may be useful for the magnetic trapping of the $^{83}$Kr* atoms.

The ground state of Krypton atom has $4s^24p^6$ configuration. The first excited state $4p^55s[3/2]_2$ is a metastable state with lifetime of ~ 40 seconds and is ~10 eV separated from the ground state. The spectroscopy can be performed by optical excitation from this metastable state to the higher state $4p^55p[5/2]_3$ using a ~ 811.5 nm laser source. The transition $4p^55s[3/2]_2$ to $4p^55p[5/2]_3$ has natural linewidth, Γ ~ 5.56 MHz.

The even isotopes of Kr such as $^{84}$Kr has nuclear spin I = 0 and has only one possible fine transition i.e. from metastable state J = 2 to excited state J' = 3. In contrast, for odd isotopes of Krypton such as $^{83}$Kr with nuclear spin I = 9/2, there are 15 possible hyperfine transitions exists in $4p^55s[3/2]_2$ to $4p^55p[5/2]_3$ manifold (see Fig. 1) [14]. The rich hyperfine structure of $^{83}$Kr* provides a number of DTLSs to realize EIA and EIT effects. In the in present work, EIT and EIA have been observed in DTLSs in $^{83}$Kr* for several hyperfine transitions ( e.g. F - F' as 7/2 – 7/2 , 9/2 - 11/2, 11/2 - 11/2 and 13/2 - 15/2).

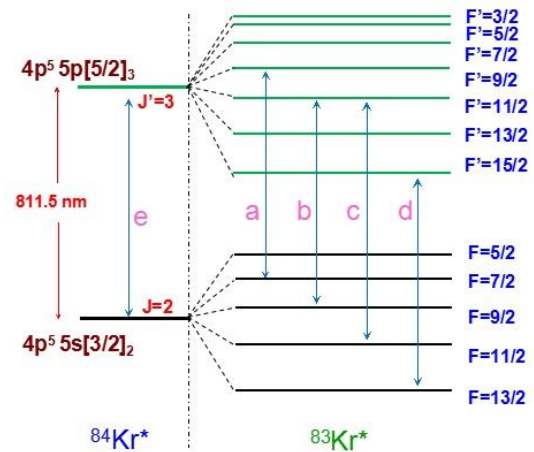

Fig.1: The relevant energy levels in $4p^55s[3/2]_2$ to $4p^55p[5/2]_3$ manifolds of metastable Krypton (Kr*). The transition 'a-d' are utilized for Degenerate Two Level Schemes (DTLSs) in $^{83}$Kr* atoms and transition 'e' is utilized for DTLS in $^{84}$Kr*.

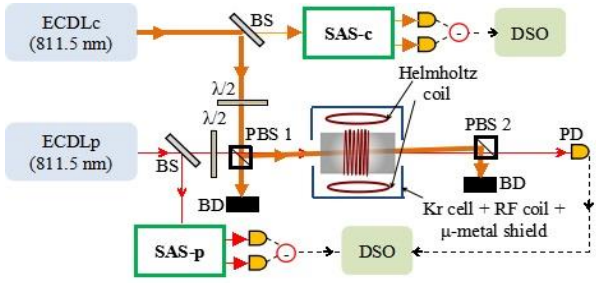

Fig.2: The schematics of the experimental setup. ECDLc(p): control(probe) laser, SAS-c(p): saturated absorption spectroscopic setups for control (probe) laser, λ/2: half waveplate, PBS: polarizing beam splitter, M: mirror, DSO: digital storage oscilloscope, PD: photodetector, BS: Beam splitter, BD: beam dump.

The schematic of the experimental setup used for this work is shown in Fig. 2. Two external cavity diode laser (ECDL) systems (Toptica, Germany) with wavelength ∼ 811.5 nm and linewidth ∼ 700 kHz have been used as control and probe lasers in the experiment. The SAS-c and SAS-p are the saturated absorption spectroscopy setups used to generate the frequency references for the control and the probe lasers and are similar to those used in our earlier work [15].

The strong control and the weak probe laser beams ($1/e^2$ spot size ∼ 0.3 mm) are in co-propagating geometry and are carefully overlapped (separated) using the polarizing beam splitters PBS1 and PBS2. This makes the plane polarized control and probe beams orthogonal to each other. Slight misalignment is introduced in the control and the probe beams to restrict the residual part of strong control beam falling on the photodetector (PD). This is also helpful to avoid beating between the probe laser beam and the residual control laser beam falling on the detector. The Kr gas cell (pressure ∼ 200 mTorr) used for the experiment is placed in between the pair of the PBSs. The gas cell is kept inside a multi-turn copper coil used for RF excitation (with frequency ∼ 30 MHz) of Kr gas. A pair of Helmholtz coil is used to produce uniform transverse magnetic field for Kr* atoms in cell. The coils and cell assembly is surrounded by a cylindrical μ-metal shield to avoid the influence of stray magnetic fields. The flux-gate magnetometer (Barington Mag-01, UK) is used (not shown in Fig. 2) for the precise measurement of applied magnetic field with the precision of 0.001 Gauss. The combinations of PBS and half waveplate (λ/2) are used to control the intensities of the control laser (power ∼ 90 mW to ∼ 239 mW) and probe laser (power ∼ 30 μW throughout the experiment) beams passing through the Kr gas cell. In our experiments, the EIA/EIT signals are obtained by measuring the variation in the transmitted probe beam signal when the control laser frequency is fixed at a resolved transition (known from the SAS-c spectrum) and the probe laser frequency is scanned around the same transition. The SAS-p signal is simultaneously recorded for the purpose of frequency reference for the probe laser. We note here that the Kr gas cells and the RF coils used in both the SAS setups are identical to that are used for recording the EIA/EIT signals.

The recorded signals of EIA and EIT effects are shown in Fig. 3. All the results shown in Fig 3 are obtained with constant control laser power of ∼ 176 mW. The EIA signals have been observed in $^{83}$Kr* in DTLSs hyperfine transitions with condition F' = F for transition 11/2 - 11/2, and with condition F' =F+1 for transition 13/2 - 15/2 (see Fig 3(C), curve (b) and curve (c)). The EIT signals have been observed with condition F' = F for the transition 7/2 - 7/2 and with condition F' = F+1 for transition 9/2 - 11/2 (see Fig 3 (A), curve (b) and curve (c)). The EIA signal obtained for open transition 11/2 - 11/2 with branching ratio, b = 0.41 [14] is larger compared to that obtained for closed transition 13/2 - 15/2 with branching ratio, b = 1. These experimental results which clearly show that the EIA can be observed for both the conditions F' = F and F' = F+1 are similar to those reported in refs. [7, 8, 11]. In addition to this, we have also observed EIT (See Fig. 3(B), curve (b)) in $^{84}$Kr* for DTLS formed by the fine transition J = 2 to J' = 3  (transition labeled as 'e' in Fig. 1). This result is similar to that reported on EIT effect recently in the metastable Neon gas atoms with condition J' = J + 1 [16].

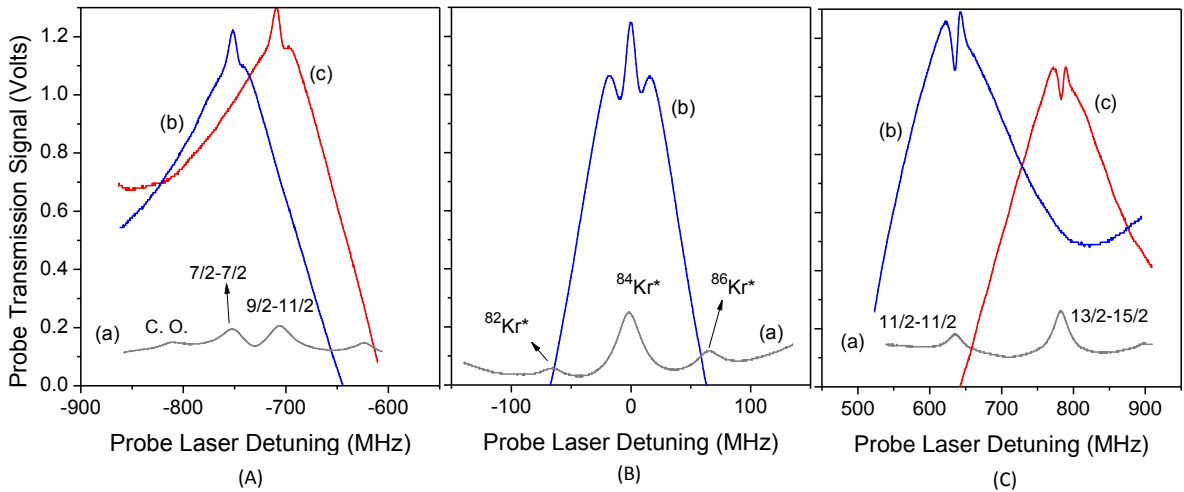

Fig.3: Observed quantum interference phenomena in Degenerate Two Level Schemes (DTLSs) of metastable Kr atoms. (A) EIT signals for 7/2 -7/2 transition (curve (b)) and 9/2 - 11/2 transition (curve (c)) of metastable $^{83}$Kr, (B) EIT signal for $^{84}$Kr transition (curve (b)) and (C) EIA signals for 11/2 -11/2 transition (curve (b)) and 13/2 - 15/2 transition (curve (c))of metastable $^{83}$Kr. Curve (a) represent simultaneously recorded corresponding SAS signals in which c.o. is the crossover peak. Here the probe laser detuning is w.r.t. $^{84}$Kr transition.

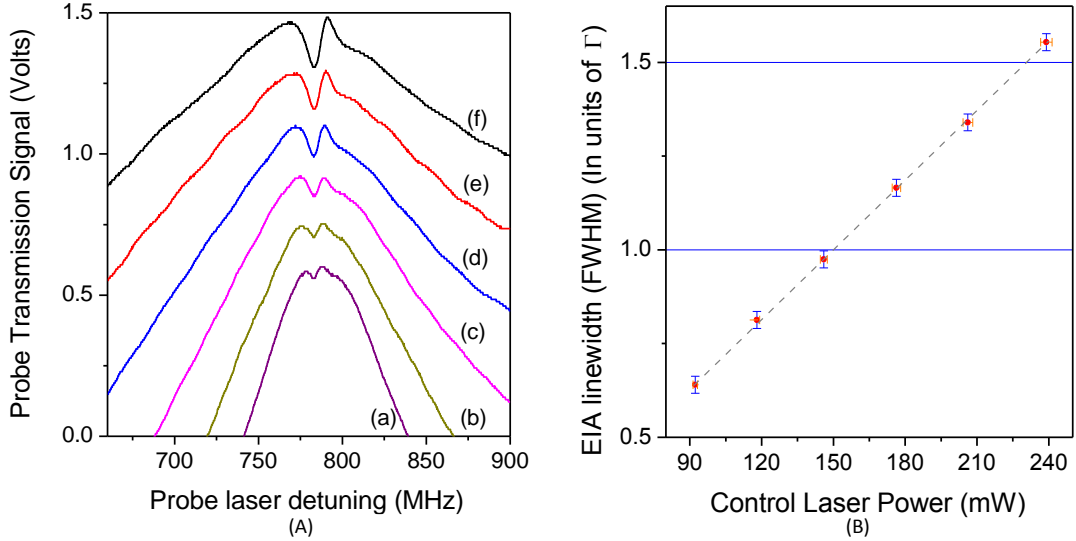

Fig.4: (A) EIA signal obtained in DTLS of ⁸³Kr* with closed (13/2 - 15/2) transition for the control laser power values of (a) 92 mW, (b) 118 mW, (c) 146 mW, (d) 176 mW, (e) 206 mW and (f) 239 mW. (B) The linewidth (FWHM) of EIA signals as a function of control laser power. The dashed line is guide to eye.

We have also investigated the dependence of narrow EIA signal of the closed transition 13/2 - 15/2 on the control laser power. The linewidth (FWHM) of each of the EIA signal is measured for the different values of control laser power used in the range of (~ 92 mW to ~ 239 mW). It is clearly observed that the linewidth of EIA signal varies linearly with the change in power of the control laser (See Fig. 4(A) and (B)). The sub-natural linewidth (FWHM) of EIA signal is obtained for the control laser intensity less than ~ 150 mW. But, for smaller intensities of control laser (< 90 mW) the amplitude of EIA signal also reduces significantly. The linewidths of EIA/EIT signals can further be narrowed by using a single laser for control and probe beam and scanning the probe beam frequency by Acousto-optical Modulator [1, 17] and further locking the single laser to a high finesse Fabry-Pérot Cavity [9].

The measurement of $g_F$ factor for the lower energy level $F_g$ of the closed transition is important for magnetic trapping of the atoms. The conventional methods for the g-factor measurement include level-crossing method [18, 19], optical pumping and magnetic resonance method [20, 21], microwave Zeeman absorption spectroscopy [22] and projectile coulomb excitation method combined with transient magnetic fields [23]. In case of noble gas atoms, g-factor measurements has been carried out in Kr by saturated absorption spectroscopy [24] and by the coulomb excitation method [25], and in Ar by opto-galvanic spectroscopy [26]. These methods [19-21] have also been used for Xe and Ne atoms. In contrast to other methods, our proposed method is entirely based on quantum interference phenomena. Using our method here, the $g_F$-factor has been measured by measuring the splitting of EIA peaks due to applied magnetic field. We used low magnetic field (< 5 Gauss), whereas other methods require hundreds of gauss magnetic field. The use of high magnetic field is likely to give error in $g_F$ measurement due to higher order (non-linear) terms in Zeeman shift.

The Landé $g_F$-factor for a hyperfine level with total momentum quantum number F can be theoretically calculated as,

$$g_{F(theoretical)} = g_J \frac{F(F+1) - I(I+1) + J(J+1)}{2F(F+1)} + g_I K \frac{F(F+1) + I(I+1) - J(J+1)}{2F(F+1)}, \quad (1)$$

where

$$g_J = g_L \frac{J(J+1) - S(S+1) + L(L+1)}{2J(J+1)} + g_S \frac{J(J+1) + S(S+1) - L(L+1)}{2J(J+1)},$$

and K = $m_e/m_p$, with $m_{e(p)}$ is mass of electron (proton). Here, L is the total orbital momentum quantum number, S is the total spin quantum number, I is the nuclear spin quantum number and $g_L$, $g_S$ and $g_I$ are Landé g-factors correspond to L, S and I respectively. The $g_F$ can be experimentally measured by the frequency separation between EIA peaks due to a applied magnetic field perpendicular to the direction of propagation of orthogonal linearly polarized control and probe beams. In the presence of magnetic field, the splitting in EIA peaks is $2\Delta$, where $\Delta$ is given as $\Delta = g_F \mu_B B / \hbar$ [1, 27].

Here $\mu_B$ is Bohr magneton, $\hbar$ is reduced Planck's constant. This gives us

$$g_{F(exp)} = \hbar \Delta / \mu_B B, \quad (2)$$

The narrow EIA signals obtained in the DTLS formed by the closed hyperfine transition 13/2 - 15/2 of 83Kr* have been used to measure $g_F$ of the lower state hyperfine state F = 13/2. The Fig. 5 shows the splitting of EIA signal in the presence of applied magnetic field B which was varied in the range from ~ 1.2 G to ~4.6 G. Fig 6(A) shows the measured variation in splitting (2Δ) in EIA peaks with the applied magnetic B. This linearity of graph ensures that the experiments are performed in the regime of linear Zeeman splitting. Using these values of Δ, the values of $g_F$ have been estimated using equation (2). Fig 6 (B) shows the variation in the measured values of $g_F$ (discrete data points) with B. Here in Fig. 6 (B), the solid line shows the average value of the measured gF and dashed line shows theoretical value of gF obtained from equation (1). For $g_{F(theoretical)}$ in equation (1), we used F = 13/2, I = 9/2, J = 2 and K = me/mp, with me(p) is mass of electron (proton). The values of constants used in Eq. (1) and (2) are taken from "CODATA recommended values of fundamental physical constants 2010" [28]. From the measured data, the average value of $g_{F(exp)}$ is 0.482(2), where number in the parenthesis shows the uncertainty (standard deviation) in the last digit. It is evident that the average value of $g_{F(exp)}$ is close to $g_{F(theoretical)}$ = 0.483.

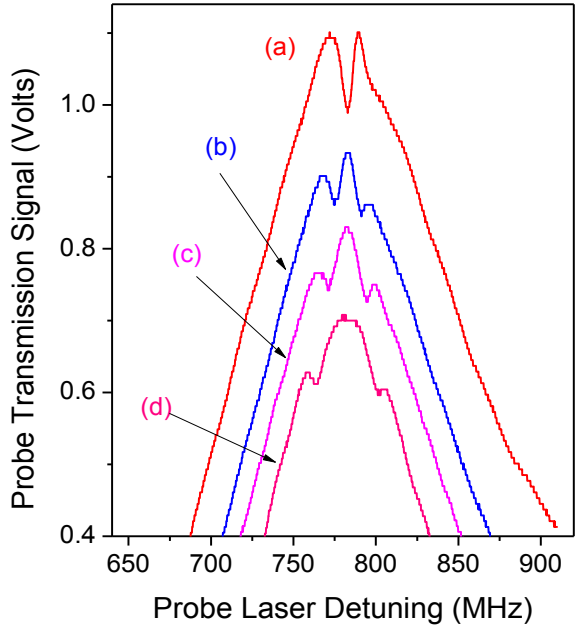

Fig.5: Observed EIA signals in the presence of the applied transverse magnetic field (B), when control and probe beams are orthogonal plane polarized. The B values are : (a) 0 G, (b) 2.045 G, (c) 2.774 G and (d) *4.594 G*. The B values are measured using fluxgate magnetometer with 0.1 % accuracy.

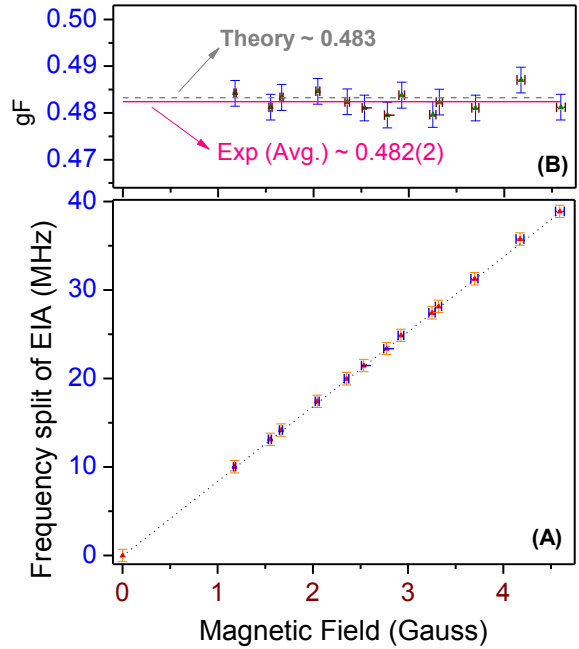

Fig.6: (A) Variation in splitting of EIA signals with the applied magnetic field. (B) Experimentally measured values of $g_{F(exp)}$. The average of $g_{F(exp)}$ is shown by solid line. The theoretical value $g_{F(theorical)}$ is shown by dashed line.

In conclusion, we have observed the narrow EIA signals in DTLS in metastable $^{83}$Kr atoms. These narrow EIA signals have been used to measure the value of $g_F$ accurately for F=13/2 hyperfine level in the $4p^55s[3/2]_2$ manifold of the $^{83}$Kr*. Since this method requires a small magnetic field to measure the $g_F$, it can be useful for measurement of the $g_F$ in other atoms also.

We thank R. S. Shinde, RRCAT, Indore for providing sensitive magnetometer for the experiments.